\documentclass[intlimits,twoside,a4paper]{article}
\usepackage[cp1251]{inputenc}

\usepackage[eqsecnum]{cmpj3}
\usepackage{bm}

\issue{2019}{22}{3}{33002}
\doinumber{10.5488/CMP.22.33002}

\title
{Continuous Aharonov--Bohm effect}

\author{V.A. Uzunova}
\address{Institute of Physics of the National Academy of Sciences of Ukraine, 46 Nauky Ave., 03039 Kyiv, Ukraine}
 
\date{Received	May 24, 2019, in final form July 18, 2019}

\begin{document}

\maketitle

\begin{abstract}
The Aharonov--Bohm effect in a model system described by the
generalized Schr\"{o}dinger equation is considered. The scattering
cross section is calculated in the standard formulation: an electron
beam impinges on a long impenetrable solenoid encompassing a
magnetic field. It is shown that the incident wave is scattered
regardless of whether the magnetic flux through the solenoid is an
integer of flux quanta whereby the scattering cross section becomes
a continuously nonzero function of the total magnetic flux. The
problem may find its application in investigations of the
soliton-magnon interaction in the low-dimensional magnetism.
\pacs 03.65-w, 03.65.Nk, 03.65.Vf
\end{abstract}

\section{Introduction}
 The Aharonov--Bohm effect demonstrates
the significance of the vector potential, $\mathbf{A}$, in the
quantum theory. It was first described by Aharonov and Bohm in
their original work \cite{AB} in 1959 and was intensively studied
afterwards by a number of authors
\cite{Brown1,Brown2,Kretzschmar,Olariu,Kobe,ShekaPRA}. The effect
refers to the interference phenomena of charged particles
(electrons) traveling around an extremely long isolated solenoid
whose radius tends to zero. The solenoid is a source of a
magnetic field, $\mathbf{B}=\textrm{rot}~\mathbf{A}$. The magnetic
flux density is concentrated in the small region inside the
solenoid, so the total flux is
$\Phi_\textrm{s}=\int\mathbf{B}\textrm{d}\mathbf{s}=\oint\mathbf{A}\textrm{d}\mathbf{l}$,
where integration can be made through every closed circuit
encircling the singularity. The movement of the particles takes place
outside the magnetic field region, where $\mathbf{B}=0$.
Nevertheless, the vector potential does have a physical effect on
their scattering. In the simplest case, when an electron moves on a
closed trajectory including the solenoid, the phase of the electron
wave function is additionally changed by $-2\piup\alpha=2\piup
\Phi_\textrm{s}/\Phi_0$, where $\Phi_0$ is the flux quantum. The
requirement for the wave function to be single-valued leads to the
quantization of the flux by $\Phi_0$. The value of $\alpha$ is
determined by a number of flux quanta in the total flux through the
solenoid and depends only on the topological invariant
$\oint\mathbf{A}\textrm{d}\mathbf{l}$. Within the problem of the
interference of two coherent electron beams bypassing the solenoid
from different sides, the vector potential manifests itself as an
additional shift of the interference picture. When the flux inside
the solenoid is a multiple of flux quanta, $\alpha$ is an integer
and the interference is constructive (the relative phase shift
depends only on the difference between the paths traveled). Noninteger
$\alpha$ means that the electron wave function acquires an
additional phase factor and the interference pattern shifts. In
fact, the solenoid plays a role of an impenetrable flux line that
makes the field-free region of space multiply connected. The interaction of particles with the topological peculiarity produces an
additional phase shift in their wave function. It can be experimentally detected by measuring the arrangement and the sharpness of the
interference bands.

Nowadays, the relevance of the problem is maintained by the development
of nanoelectronic, mesoscopic and spintronic devices, and quantum
computers. The Aharonov--Bohm effect found its application in
single-electron transistors with suspended nanotubes
\cite{Rastelli,Skorobagatko}. Electrons passing through the device
interfere on the nanotubes vibrating in the presence of a magnetic
field \cite{Shekhter}. The Aharonov--Bohm effect is often involved in
comprehending different phenomena in the quantum physics. The existence
of an analogous effect was discussed in the physics of
low-dimensional magnetism: in the problems of vortex-magnon
scattering in easy-plane ferromagnets \cite{Mertens,Sheka} and
skyrmion-magnon scattering \cite{ShekaMer01,Nagaosa}, including
skyrmions in chiral magnets \cite{Nagaosa}. The non-trivial topology
of an isolated soliton gives rise to terms in the magnon dynamical
equations similar to the term with vector potential $\mathbf{A}$ in
the Schr\"{o}dinger equation.
 In cone state ferromagnets, this Aharonov--Bohm type of interaction with the nonlinear vortex
core causes a significant doublet splitting of magnon modes
\cite{IvWysinPRB02,Uzunova19}. In chiral magnets, there appears a skew and rainbow
scattering of magnons, characterized by an asymmetric
differential cross section \cite{Garst,GarstFNT}. However, equations
describing magnon dynamics can be much more complicated than the
standard Schr\"{o}dinger equation. For this reason, various
generalizations of the Aharonov--Bohm effect are of great interest.

 In this paper, we investigate the existence
of the Aharonov--Bohm effect in a model system that is described by
the generalized Schr\"{o}dinger equation. It appears in the study of
quantum many-particle systems by means of the approximating
Hamiltonian method originally developed by N.N. Bogolyubov.
The method is equivalent to introducing some quasiparticles that
are capable of condensing into a single quantum state \cite{deGennes,Landau9}.
In the effective Hamiltonian, this leads to the appearance of
anomalous average terms describing the exchange of particles between
condensate and noncondensate states. The model approach allows one
to turn to the exactly solvable quantum models that cannot be
obtained by means of the perturbation theory and, therefore, occupies
an important place in condensed matter physics \cite{Bogolyubov}.
However, the Aharonov--Bohm problem in such systems, to the best of
our knowledge, has not been considered yet.

In the present work, we obtain an asymptotic solution of the
generalized Schr\"{o}dinger equation in presence of the vector
potential (section \ref{sect2}) and solve a direct scattering
problem (section \ref{sect3}). We numerically calculate the
scattering cross section and show that the scattering picture is
qualitatively different from the one of the original Aharonov--Bohm
problem (section \ref{sect4}). We also briefly discuss possible
applications of the model problem in section \ref{concl}.

\section{Solution of the generalized Schr\"{o}dinger equation in the presence of a magnetic flux
line}\label{sect2}
 The generalized
Schr\"{o}dinger equation may be derived as the Euler equation of
some variational problem \cite{Bogolyubov}, in which the Lagrangian
density has the form
 \begin{equation}\label{L}
L=\ri\hbar\psi^* \frac{\partial\psi}{\partial t} -\psi^*
\frac{1}{2m_0}
\left(-\ri\hbar\nabla-\frac{e}{c}\mathbf{A}\right)^2\psi -\psi^*
a\psi-\frac{b}{2}(\psi^*\psi^*+\psi\psi).
\end{equation}
For convenience, we use the original formulation of the
Aharonov--Bohm problem as a concrete example that allows one to rely on
the formulae obtained in the work \cite{AB} without changing them.
Additionally, we introduce coordinate independent normal and abnormal
pairing potentials indicated by $a$ and $b$, respectively. Specific
values of these potentials, as well as a possible origin of the
Lagrangian density~(\ref{L}), are discussed in section \ref{concl}.

\begin{figure}[!t]
\centerline{\includegraphics[width = 0.6\textwidth]{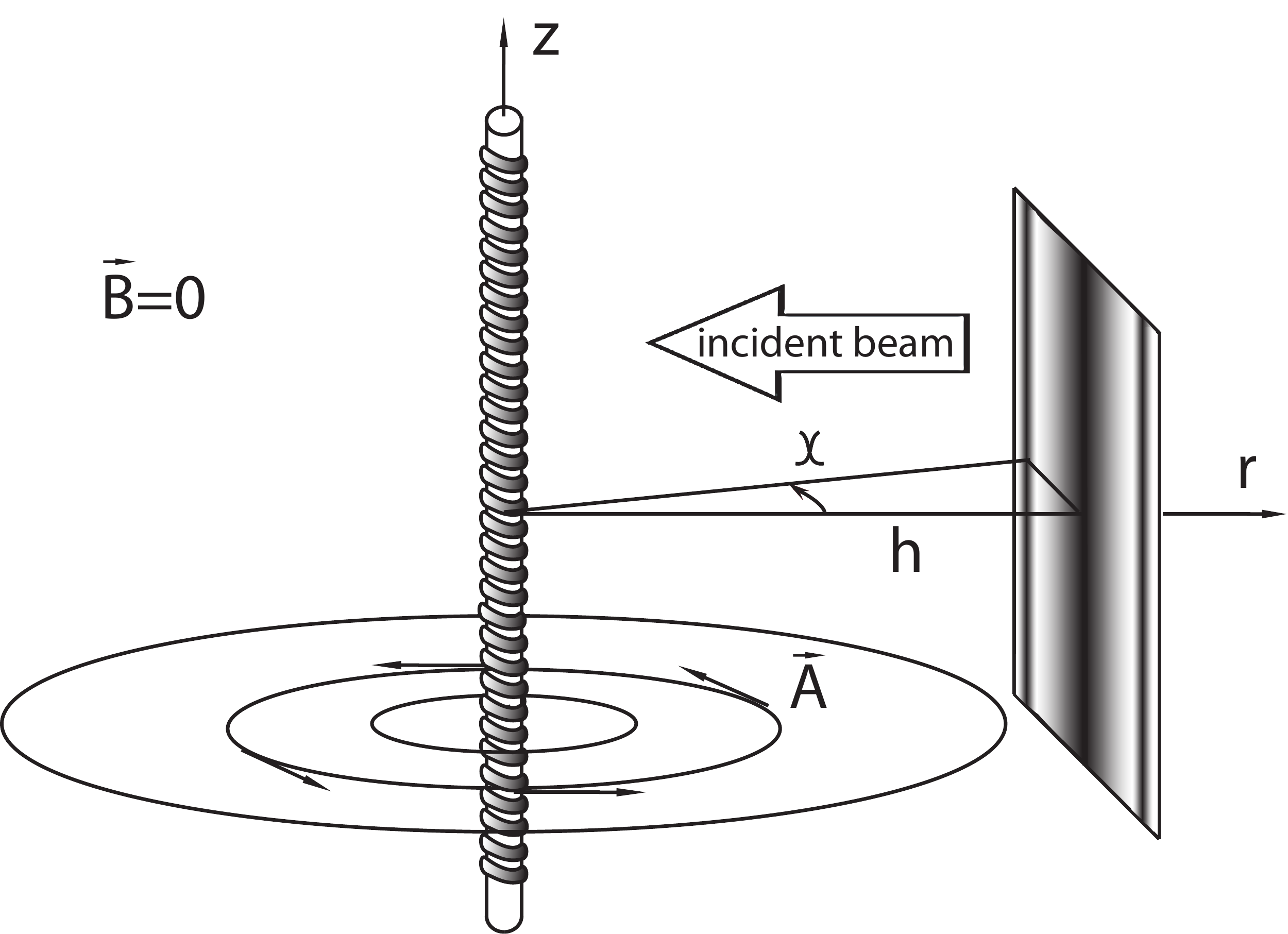}}
\caption{The schematic picture of the system: a coherent electron
beam is scattered by an infinitely long impenetrable solenoid. The
incident beam is coming from the right where $\chi=0$. Magnetic
field outside the solenoid: $\mathbf{B}=0$. Intensity of bands on a
screen is determined by the scattering cross section shown in figure
\ref{Fig2}.}\label{Fig3}
\end{figure}
A schematic picture of the system is shown in figure \ref{Fig3}. A
coherent electron beam is scattered by an infinitely long
impenetrable solenoid located along the $z$-axis of the cylindrical
geometry. Its radius tends to zero, while the total flux,
$\Phi_\textrm{s}$, remains fixed. The vector potential of the
solenoid is $\mathbf{A}=\Phi_\textrm{s}/2\piup r\mathbf{e}_\chi$,
where $\mathbf{e}_\chi$ is the unit azimuthal vector, $r$ and $\chi$
are the radial and azimuthal coordinates, respectively.
 We also
use the traditional notation of the flux quantum $ \Phi_0= 2\pi\hbar c/e$,
where $c$ is the speed of light, $e$ is the electron charge,
$\hbar$ is the Planck constant. The energy and the wave vector
of the electron in the incident wave are denoted as $\omega\hbar$
and $\mathbf{k}$, and $m_0$ is the electron mass.

The variation of the Lagrangian $\mathcal{L}=\int L
\textrm{d}\textbf{r}$ results in
 \begin{equation}\label{Shr}
 \ri\hbar\frac{\partial\psi}{\partial
t}=\frac{1}{2m_0}\left(-\ri\hbar\nabla-\frac{e}{c}\mathbf{A}\right)^2\psi+a\psi+b\psi^*.
\end{equation}

The axial symmetry of the problem means that the solution of
equation~(\ref{Shr}) does not depend on $z$-coordinate. We separate
azimuthal and radial variables by writing the electron wave function
as the sum of partial waves
 \begin{equation}
\psi=\sum_m \left[A_m u(r)\exp{(\ri m\chi-\ri\omega
t)}\vphantom{\sum}+B_m v(r)\exp(-\ri m\chi+\ri\omega t)\right].
 \end{equation}
 They are marked with azimuthal numbers, $m=0,
\pm1,\pm2\ldots$, and have amplitudes $A_m$ and $B_m$. Radial
functions $u(r)$ and $v(r)$ depend on $m$ and can be expressed
through some cylindrical
 functions. Thus, for each pair of the radial functions, an eigenvalue problem can be
 formulated as follows:
\begin{eqnarray}\label{uv}
\begin{Vmatrix}
\hat{H}_+ & b \\ -b & -\hat{H}_-
\end{Vmatrix}
\begin{Vmatrix}
u\\ v
\end{Vmatrix}
=\omega\hbar
\begin{Vmatrix}
u\\ v
\end{Vmatrix},
\end{eqnarray}
where
\begin{equation}
\hat{H}_{\pm}=-\frac{\hbar^2}{2m_0}\left[\frac{\textrm{d}^2}{\textrm{d}
r^2}+\frac{\textrm{d}}{r\textrm{d}
r}-\frac{(m\pm\alpha)^2}{r^2}\right]+a.
\end{equation}
 Note that problem~(\ref{uv}) is invariant under the conjugations
$\omega\hbar\rightarrow -\omega\hbar$, $m\rightarrow -m$, and $u
\rightarrow v$. Thus, without loss of generality, we can restrict
ourselves to a case of the positive energy only, $\omega\hbar>0$.

If there is no magnetic flux $\alpha=0$, then positive and negative
energy solutions of eigenvalue problem~(\ref{uv}) can be easily
separated by the Bogolyubov transformation
\begin{eqnarray}\label{BT}
\begin{Vmatrix}
\tilde{u}\\ \tilde{v}
\end{Vmatrix}=R
\begin{Vmatrix}
{u}\\ {v}
\end{Vmatrix}
,\;\; R=
\begin{Vmatrix}
\cos\varepsilon & -\sin\varepsilon\\ \sin\varepsilon &
\cos\varepsilon
\end{Vmatrix}
\end{eqnarray}
with the rotation angle
\begin{eqnarray}
\tan 2\varepsilon=\frac{b}{\omega\hbar}.
\end{eqnarray}
 The result can be interpreted as the occurrence of certain
quasiparticles \cite{Bogolyubov} with the energy
\begin{equation}
\Lambda=\sqrt{b^2+\omega^2\hbar^2}
\end{equation}
and a gap spectrum
\begin{equation}\label{dl}
\Lambda=a+\frac{\hbar^2  k^2}{2m_0}.
\end{equation}
In terms of $\omega\hbar$, it is
\begin{equation}\label{dlo}
\omega^2\hbar^2=a^2-b^2+\frac{\hbar^2  k^2}{m_0}a+\frac{\hbar^4
k^4}{4m_0^2}.
\end{equation}

In the most general case of a nonzero $\Phi_\textrm{s}$, the analysis
of eigenvalue problem~(\ref{uv}) is much complicated by the
fact that there are two different equations for $u$ and $v$. The
Bogolyubov transformation gives
\begin{eqnarray}\label{uvtilda}
\left(
\begin{Vmatrix}
\hat{H}_0 & 0 \\ 0 & -\hat{H}_0
\end{Vmatrix}
+\hat{X}R^{-2} \right)
\begin{Vmatrix}
\tilde{u}\\ \tilde{v}
\end{Vmatrix}
=\Lambda
\begin{Vmatrix}
\tilde{u}\\ \tilde{v}
\end{Vmatrix},
\end{eqnarray}
where
\begin{equation}
\hat{H}_0=-\frac{\hbar^2}{2m_0}\left(\frac{\textrm{d}^2}{\textrm{d}
r^2}+\frac{\textrm{d}}{r\textrm{d}
r}-\frac{m^2+\alpha^2}{r^2}\right)+a\,,\;\;
\hat{X}=\frac{\hbar^2}{2m_0}\frac{2m\alpha}{r^2}.
\end{equation}
A new pair of radial eigenfunctions $\tilde{u}$ and $\tilde{v}$ are
related due to the potential $\hat{X}$ and can be analyzed only in the
sense of their asymptotic behavior. Such an approximation is valid
since $\hat{X}$ strongly decreases with the distance. As we are
interested only in positive energy solutions ($\Lambda>0$), the
function $\tilde{u}$ takes a role of a master function, while
$\tilde{v}$ becomes a slave one.

Following the idea of the appearance of quasiparticles, we regard
$\tilde{u}(r)$ as the radial part of a new wave function
\begin{equation}\label{psitilde}
\tilde{\psi}=\sum_mc_m\tilde{u}(r)\exp\left(\ri m\chi-\ri
\frac{\Lambda}{\hbar} t\right).
\end{equation}
It describes a particle with energy $\Lambda$ and is the solution of
an effective Schr\"{o}dinger equation
 \begin{equation}\label{ShrPsi}
 \ri\hbar\frac{\partial\tilde{\psi}}{\partial
t}=\frac{1}{2m_0}\left(-\ri\hbar\nabla-\frac{e}{c}\tilde{\mathbf{A}}\right)^2\tilde{\psi}+V\tilde{\psi}.
\end{equation}
Here, $\tilde{\mathbf{A}}=\mathbf{{A}}\cos
2\varepsilon=-\tilde{\alpha}\mathbf{e}_\chi c\hbar/er$ and
 \begin{equation}\label{}
V=\frac{\hbar^2}{2m_0}\frac{\tilde{\alpha}^2\tan^22\varepsilon}{r^2}+a.
\end{equation}
The effective potential $V$ has the radially symmetric $r$-dependent
part and the constant part $a$. The last one can be considered as an
energy level relative to which the quasiparticle energy is measured.
Thus, the incident quasiparticles are scattered on the effective
barrier $\sim 1/r^2$ and the total flux is renormalized as
$\tilde{\Phi}_\textrm{s}=\Phi_\textrm{s}\cos 2\varepsilon$.

The radial function $\tilde{u}(r)$ from the formula (\ref{psitilde})
satisfies the Bessel equation
\begin{equation}\label{Bessel}
\frac{\hbar^2}{2m_0}\left(\frac{\textrm{d}^2}{\textrm{d}
r^2}+\frac{\textrm{d}}{r\textrm{d}
r}-\frac{\nu^2}{r^2}\right)\tilde{u}=\Lambda\tilde{u}
\end{equation}
 with the generally noninteger index
\begin{equation}\label{nu}
\nu^2=(m+\tilde{\alpha})^2+\tilde{\alpha}^2\tan^22\varepsilon.
\end{equation}
On the formulation of the problem, the particles do not penetrate
into the region of close vicinity of the solenoid and the
probability of finding the particle at the origin approaches zero.
For this reason, we restrict a general solution of
equation~(\ref{Bessel}) to the Bessel function $J_\nu(kr)$ of the
positive index only, $\nu>0$. The spectrum of the particle is given
by expression~(\ref{dl}).

\section{Scattering problem for quasiparticles}\label{sect3}
Taking into account the results of the previous section, the
scattering problem for quasiparticles of the energy $\Lambda$ and
impulse $\hbar \mathbf{k}$ can be formulated in a canonical way. A
steady beam of particles is scattered by an impenetrable magnetic
flux line with the total flux $\tilde{\Phi}_\textrm{s}$. The
effective potential $V$ created by the flux line is radially
symmetric and short ranged. Hence, one can distinguish between incident and scattered waves at large values of $kr$.

The total wave function $\tilde{\psi}$ is given by
expression~(\ref{psitilde}). The coefficient $c_m$ is chosen in a
way to present $\tilde{\psi}$ as a superposition of incident plane
wave $\tilde{\psi}_{\textrm{inc}}$ and the scattered cylindrical
wave,
\begin{equation}\label{scat}
\tilde{\psi}=\tilde{\psi}_{\textrm{inc}}+\mathcal{F}(\chi)\frac{1}{\sqrt{kr}}\exp\left(\ri
kr-\ri \frac{\Lambda}{\hbar}t\right).
\end{equation}
Scattering function $\mathcal{F}(\chi)$ describes directions of
 strong and weak scattering of particles. It is fully determined by
$\tilde{\alpha}=\alpha\cos 2\varepsilon$ which is exactly the number
of flux quanta in the total flux $\tilde{\Phi}_\textrm{s}$ taken
with the opposite sign.
 The incident
wave is coming from the right, where $\chi=0$,
\begin{equation}\label{inc}
\tilde{\psi}_{\textrm{inc}}=\exp\left(-\ri
kr\cos\chi-\ri\frac{\Lambda}{\hbar}t -\ri\tilde{\alpha}\chi\right).
\end{equation}
Initial phase $\tilde{\alpha}\chi$ is chosen to conserve the current
density,
 \begin{equation}\label{}
\mathbf{j}=\frac{\hbar
}{2m_0\ri}\left(\tilde{\psi}^*\nabla\tilde{\psi}-\tilde{\psi}\nabla\tilde{\psi}^*\right)-\frac{e}{m_0c}\tilde{\mathbf{A}}\tilde{\psi}^*\tilde{\psi},
\end{equation}
that takes into account the renormalized vector potential
$\tilde{\mathbf{A}}$. In order to decompose the incident wave, we use
the well-known equality expressing a plane wave through the sum of
partial cylindrical waves,
\begin{equation}\label{plane}
\exp(-\ri kr\cos\chi)=\sum_m(-\ri)^mJ_m(kr)\exp{(\ri m\chi)}.
\end{equation}

The behavior of wave function $\tilde{\psi}$ in the limit
$kr\rightarrow\infty$ is determined by the far asymptotic of the Bessel
function
\begin{equation}\label{as}
J_\nu(kr)\sim\frac{1}{\sqrt{2\piup rk}}\left[\exp\left(\ri
kr-\ri\frac{\nu\piup}{2}-\ri\frac{\piup}{4}\right)\vphantom{\frac{1}{1}}+\exp\left(-\ri
kr+\ri\frac{\nu\piup}{2}+\ri\frac{\piup}{4}\right)\right].
\end{equation}
The first exponent presents a circular wave diverging from the
center, and the second one is a circular wave converging to the
center. The correct coefficient, $c_m=(-i)^\nu$, in the wave
function~(\ref{psitilde}) corresponds to the converging circular
wave from the incident plane wave only. Finally, the wave function
is as follows:
\begin{equation}\label{Psi_tilda}
\tilde{\psi}=\sum_m (-\ri)^\nu J_\nu(kr)\exp\left(\ri m\chi-\ri
\frac{\Lambda}{\hbar}t\right).
\end{equation}
In fact, this formula is the total wave function of
the system satisfying the boundary conditions at $r=0$ and at
infinity. It is sufficient for a complete numeric analysis of the
scattering problem. Nevertheless, we try to get some analytical
results.

Let us note that assuming $V=0$ in equation~(\ref{ShrPsi}) would
turn it into the Schr\"{o}dinger equation of free particles under
the action of the vector potential $\tilde{\mathbf{A}}$. In the
formulation of the scattering problem, this equation has the solution
obtained by Aharonov and Bohm,
\begin{equation}\label{Psi_AB}
\tilde{\psi}_{\textrm{AB}}=\sum_m (-\ri)^{|m+\tilde{\alpha}|}
J_{|m+\tilde{\alpha}|}(kr)\exp \left( \ri
 m\chi-\ri \frac{\Lambda}{\hbar}t \right),
\end{equation}
for details see the article \cite{AB}. Here, the index of the Bessel
function, $|m+\tilde{\alpha}|$, denotes a phase shift of the wave
function due to the presence of the magnetic flux line. When
$\tilde{\alpha}$ is integer, $\tilde{\psi}_{\textrm{AB}}$ can be
easily converted into a plane wave by shifting the summation index,
$m'=m+\tilde{\alpha}$, and using equality~(\ref{plane}). Thus, there is
no scattering for integer $\tilde{\alpha}$.

In the wave function~(\ref{Psi_tilda}), the index of the Bessel function,
$\nu$, is given by expression~(\ref{nu}). The index is noninteger
for all $\tilde{\alpha}\neq 0$. Therefore, there is always a nonzero
phase shift that is reflected in the scattering cross section,
$\sigma=\mathcal{F}\mathcal{F}^*$. Scattering function $\mathcal{F}$
from representation (\ref{scat}) is obtained by calculating the far
asymptotic of the difference
$\tilde{\psi}-\tilde{\psi}_{\textrm{AB}}$. We use the fact that the
asymptotic expression~(\ref{as}) is correct for a Bessel function of
any index and apply it to the functions $J_\nu$ and
$J_{|m+\tilde{\alpha}|}$. Finally,
\begin{eqnarray}\label{F}
\mathcal{F}&=&\mathcal{F}_{\textrm{AB}}-\sqrt{\frac{2\ri}{\piup}}
\sum_{m=-\infty}^\infty\sin\left\{\frac{\piup}{2}\Big[\sqrt{(m+\tilde{\alpha})^2+\tilde{\alpha}^2\tan^22\varepsilon}
-|m+{\tilde{\alpha}}|\Big]\right\}\nonumber\\
&\times&\exp\left\{\ri
m\chi-\ri\frac{\piup}{2}\Big[\sqrt{(m+\tilde{\alpha})^2+\tilde{\alpha}^2\tan^22\varepsilon}
+|m+{\tilde{\alpha}}|\Big]\right\}.
\end{eqnarray}
The first term remains for the Aharonov--Bohm scattering function,
\begin{equation}\label{}
{\mathcal{F}}_{\textrm{AB}}=\frac{1}{\sqrt{2\piup\ri}}\frac{\sin(\piup{\tilde{\alpha}})}{\cos(\chi/2)}\exp(-\ri{\chi}/{2}).
\end{equation}
Its scattering cross section,
$\sigma_{\textrm{AB}}=\mathcal{F}_{\textrm{AB}}\mathcal{F}_{\textrm{AB}}^*$,
is well-known,
\begin{equation}\label{sigmaAB}
{\sigma}_{\textrm{AB}}=\frac{1}{{2\piup
}}\frac{\sin^2(\piup\tilde{\alpha})}{\cos^2(\chi/2)}.
\end{equation}
The second term in expression~(\ref{F}) corresponds to the
scattering on the potential barrier and tends to zero in the special
case of $b=0$, implying $\tan2\varepsilon=0$. It is due to the barrier that
particles are scattered by the flux line even when the number of
flux quanta is integer.

\section{Scattering cross section calculation and its comparison with the \\ original Aharonov--Bohm problem}
\label{sect4}
In order to demonstrate the difference between
the results obtained and the original effect, we set up an imaginary
experiment on the scattering of quasiparticles.
A screen sensitive to this kind of particles is placed at a distance
$h$ from the vortex line, see figure \ref{Fig3}. It is assumed that
the incident beam is uniform along the $z$-axis (slit), so the movement of
particles can be considered only in the plane $z=0$. The
theory is applicable when the size of the system $h$ is much larger
than the wavelength $\lambda=2\piup/k$ of the incident wave, i.e.,
$kh\gg 2 \piup$.
\begin{figure}[!b]
\centerline{\includegraphics[width = 0.6\textwidth]{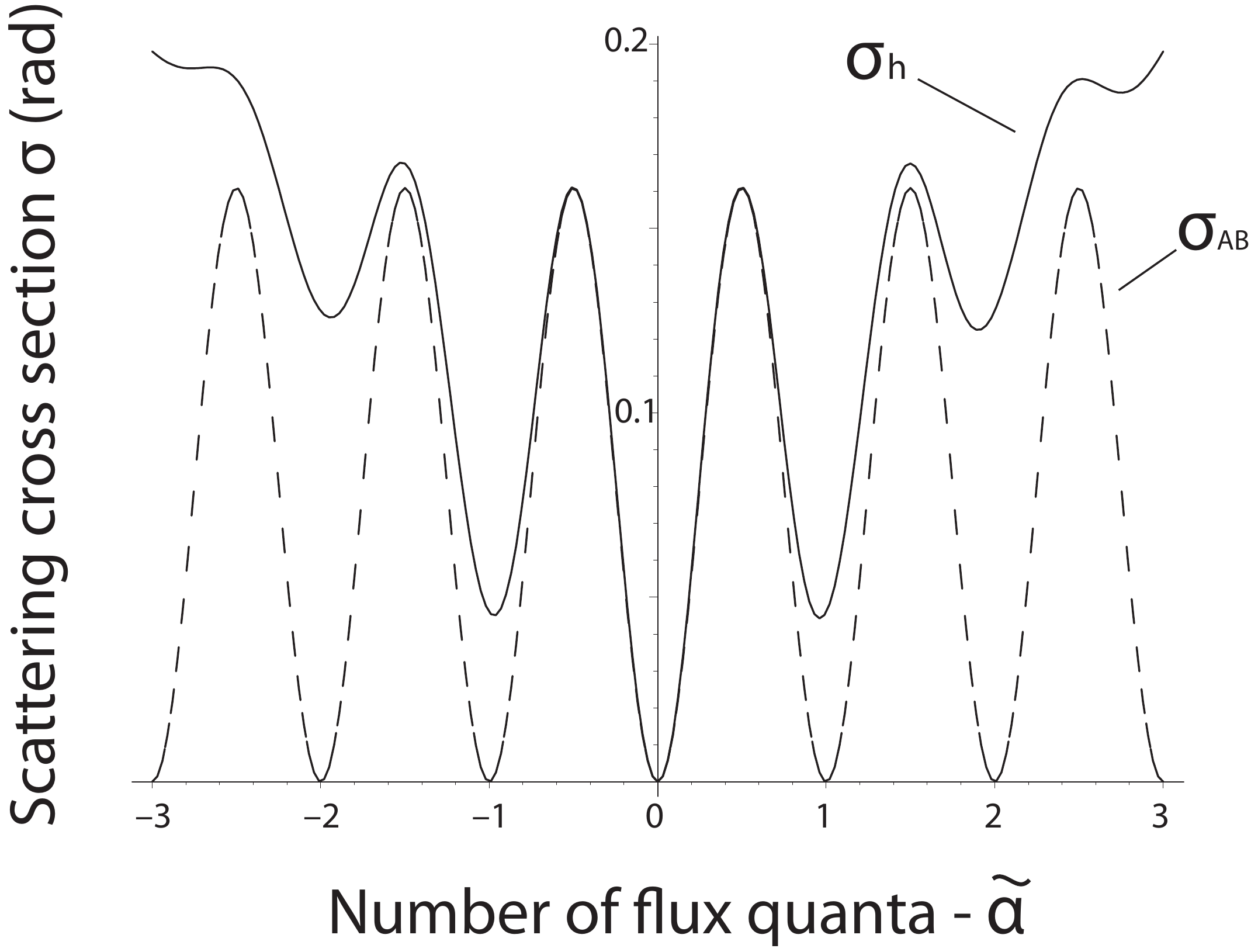}}
\caption{The scattering cross section $\sigma_{\textrm{h}}$ (solid
line) in comparison with the Aharonov--Bohm scattering cross section
${\sigma}_{\textrm{AB}}$ (dashed line) as functions of the number of
flux quanta $-\tilde{\alpha}$. For calculation we used
$b/\omega\hbar=0.2$, $kh=100$~rad, $M=900$, $\chi=\piup/15$~rad.
}\label{Fig1}
\end{figure}
\begin{figure}[!t]
\centerline{\includegraphics[width = 0.6\textwidth]{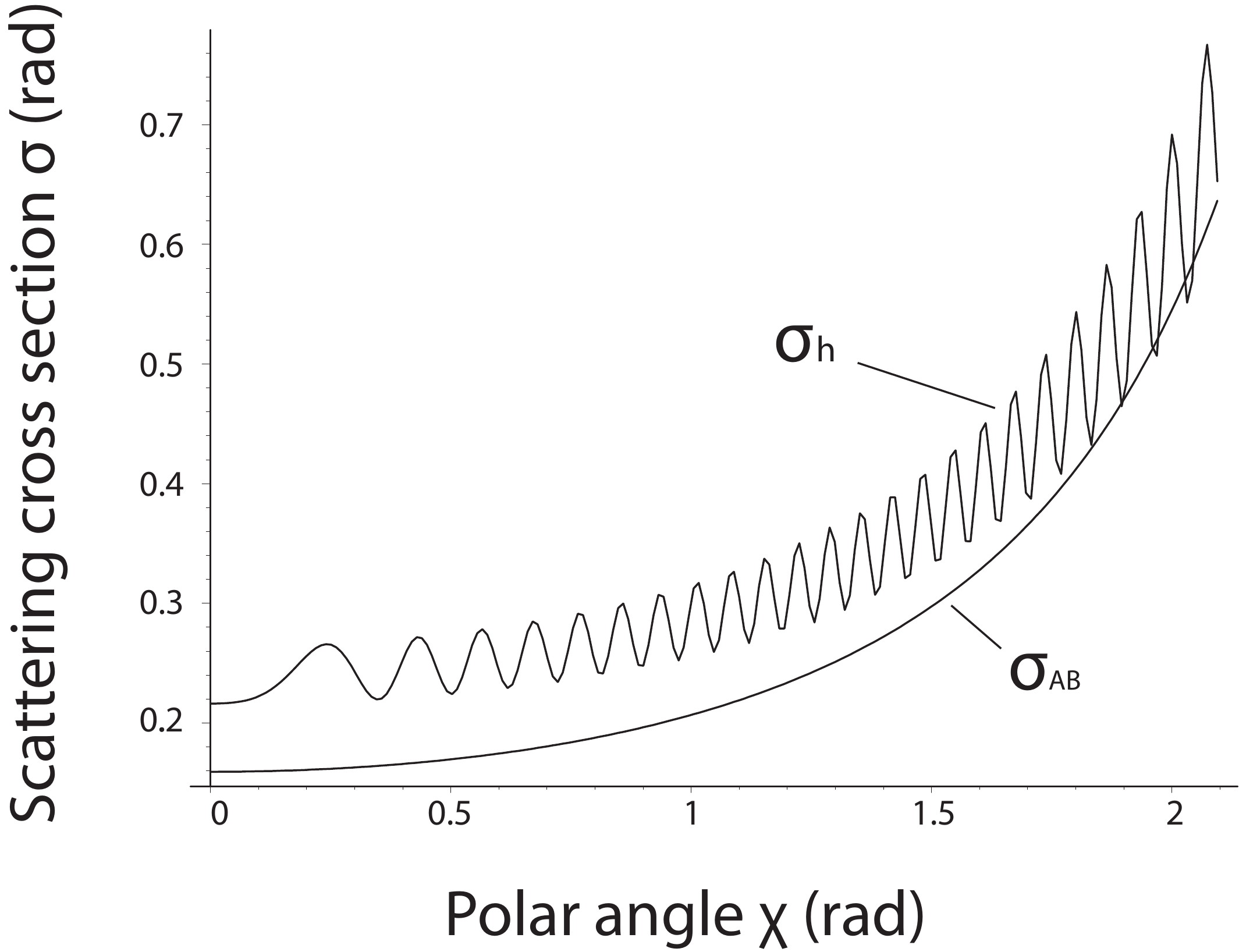}}
\caption{The angular dependence of the scattering cross section
$\sigma_{\textrm{h}}(\chi)$; $\chi=0$ indicates the direction
opposite to the incident beam. Smooth line is the Aharonov--Bohm
scattering cross section, ${\sigma}_{\textrm{AB}}(\chi)$. For
calculation we used $b/\omega\hbar=0.2$, $kh=100$~rad, $M=600$,
$\tilde{\alpha}=3.5$. }\label{Fig2}
\end{figure}

 The expression~(\ref{F}) is derived by using the asymptotics of the Bessel
 functions by type~(\ref{as}) that are valid only when the argument of the Bessel function is much larger than the index.
For this reason, the expression~(\ref{F}) is not suitable for
numerical calculations that use a finite sum instead of
infinite. Hence, we construct a formula for calculations using the exact
expressions for the wave functions, (\ref{inc}) and
(\ref{Psi_tilda}), and representation (\ref{scat}). Namely,
\begin{equation}\label{}
\mathcal{F}_{\textrm{h}}=\sqrt{kh}\exp(-\ri k h)\left[ \sum_{m=-M}^M
(-\ri)^\nu J_\nu (kh) \exp(\ri m\chi)-\exp(-\ri k h\cos\chi-\ri
\tilde{\alpha}\chi)\right],
\end{equation}
where $M$ is a maximum value of $m$ to which the numerical summation
is performed. The scattering pattern on the screen is determined by
the cross section
$\sigma_{\textrm{h}}=\mathcal{F}_{\textrm{h}}\mathcal{F}_{\textrm{h}}^*$
which is the probability of a particle to be scattered through a
given azimuthal angle $\chi$.

 Numerical results for the scattering cross section
$\sigma_{\textrm{h}}$ are given in figures~\ref{Fig1} and~\ref{Fig2}
along with ${\sigma}_{\textrm{AB}}$ for comparison.
 As it is clearly seen from figure~\ref{Fig1}, the Aharonov--Bohm scattering cross section, ${\sigma}_{\textrm{AB}}$,
vanishes for every integer $\tilde{\alpha}$ that corresponds to an
integer number of flux quanta. By contrast, ${\sigma}_{\textrm{h}}$
is nonzero in the presence of any nonzero magnetic flux through the
solenoid. In figure~\ref{Fig2} there is shown the dependence of the
scattering cross section on the azimuthal angle $\chi$. The
Aharonov--Bohm scattering gives a smooth distribution of particles on
the screen increasing from the center to the edges as ${\cos^{-2}(\chi/2)}$,
see expression (\ref{sigmaAB}). The angular dependence of
${\sigma}_{\textrm{h}}$ is much more complicated and leads to the
appearance of bands on the screen. The reason here is the different
scattering of partial cylindrical waves with different values of the
azimuthal number $m$.

 In summary, the manifestations of the Aharonov--Bohm effect in systems described by the generalized Schr\"{o}dinger equation
 loses its discreteness. The magnetic flux through the isolated region causes the scattering of particles for any nonzero value of the
total flux. In real systems the effect can be attributed to the
complex interaction with another particle system which leads to the
emergence of new channels of scattering.

\section{Conclusions}\label{concl}
To conclude, let us make some remarks on possible applications of the
model problem. Lagrangian density~(\ref{L}) appears in the study of
quantum many-particle systems with nonconserved number of particles
by reducing the problem to effective single-particle
equations. First, it is important to notice that approximating
Hamiltonian for a homogeneous electron gas in the
Bardeen--Cooper--Schrieffer theory contains a $b$-proportional term
(abnormal pairing potential) associated with the attraction between
particles \cite{deGennes}. For this term to be non-zero, it is
necessary that the electrons should be connected with the other particles,
because the usual Coulomb interaction leads only to repulsion. In
superconducting systems, the basic cause of the abnormal electron
pairing is the electron-phonon interaction. However, relevant
single-particle equations, known as the Bogolyubov equations, are
not equivalent to the generalized Schr\"{o}dinger equation
(\ref{Shr}). Therefore, the model problem considered in the article
is not applicable to the fermion systems. More promising in this
regard is the problem of the weakly nonideal Bose gas. Its
condensate wave function satisfies the Gross--Pitaevskii equation.
Linearized equations for small oscillations of the condensate wave
function are of the form (\ref{uv}) where $a=b=\mu$ is the chemical
potential \cite{Landau9}. The spectrum of the excitations coincides
with the expression (\ref{dlo}).

 Another application area is the study of spin waves in the presence of topological solitons
 in the two-dimensional model of the Heisenberg ferromagnet. The
Landau--Lifshitz equations describing the magnon dynamics on the
soliton background can be presented in the form (\ref{Shr}) with
dimensionless potentials $a$ and $b$ tending to constant values at
distances $r\gg l_0$, where $l_0$ is a characteristic length scale.
Values of $a$ and $b$ are different for each specific system: in the
case of a vortex in the easy-plane ferromagnet $a=b=1/2$, see
\cite{Sheka}; in the easy-cone state of uniaxial magnets with
comparable second-order and fourth-order anisotropy
$a=b=\sin^22\bar{\theta}/4$, where $\bar{\theta}$ is the opening
angle of the cone, see \cite{Uzunova19}. The ``mass'' $m_0$ as well as
the characteristic length $l_0$, which defines the limits of the
validity of the theory, are expressed in terms of the exchange and
anisotropy constants and the saturation magnetization. The role of
the vector potential $\mathbf{A}$ is played by the vector
$q\cos\theta_0/r\mathbf{e}_\chi$, where $q$ is the topological
charge and $\theta_0$ is the polar angle of the soliton solution.
This expression is convenient for either the vortex~\cite{Sheka} or
the Belavin--Polyakov soliton \cite{Ivanov95,IvShekaJETPL}, for the
skyrmion vector potential additionally includes a term proportional
to $\sin\theta_0$ and the Dzyaloshinskii--Moriya momentum
\cite{Garst}. For the most of the magnon modes, the wave function
strongly decreases with the distance from the soliton core, i.e.,
magnons can be regarded as being apart from the core region. Thus, the
continuous Aharonov--Bohm scattering takes place in the systems. In
the finite geometry, it results in the splitting of magnon modes.

\section{Acknowledgement}

The author is deeply grateful to B.A. Ivanov for suggesting the
problem and many helpful discussions.

\ukrainianpart

\title{Неперервний ефект Ааронова-Бома}
\author{В.О. Узунова}
\address{Інститут фізики НАН України, просп. Науки, 46, 03028 Київ, Україна}
\makeukrtitle

\begin{abstract}
\tolerance=3000%
Розглянуто ефект Ааронова-Бома в модельній системі,  що описується
узагальненим рівнянням Шредінгера. Обчислено переріз розсіяння в
стандартному формулюванні: пучок електронів падає на довгий
непроникний соленоїд з магнітним полем. Показано, що падаюча хвиля
розсіюється незалежно від того чи є потік через соленоїд цілим
числом квантів потоку, внаслідок чого переріз розсіяння стає
неперервно ненульовою функцією повного магнітного потоку. Ця
проблема може знайти своє застосування у дослідженні
солітон-магнонної взаємодії в низькорозмірному магнетизмі.

\keywords ефект Ааронова-Бома, рівняння Шредінгера, теорія розсіяння
\end{abstract}

\begin{thebibliography}{10}

\bibitem{AB} Aharonov Y., Bohm D., Phys. Rev., 1959, \textbf{115}, No.~3,
            485--491, \doi{10.1103/PhysRev.115.485}.
\bibitem{Kretzschmar} Kretzschmar M., Z. Physik, 1965, \textbf{158}, No.~1, 84--96, \doi{10.1007/BF01381305}.
\bibitem{Olariu} Olariu S., Popescu I.I., Rev. Mod. Phys., 1985, \textbf{57}, No.~2,  339--436, \doi{10.1103/RevModPhys.57.339}.
\bibitem{Brown1} Brown R.A., J. Phys. A: Math. Gen., 1985,  \textbf{18}, No.~13, 2497--2508, \doi{10.1088/0305-4470/18/13/025}.
\bibitem{Brown2} Brown R.A., J. Phys. A: Math. Gen., 1987, \textbf{20}, No.~11, 3309--3326, \doi{10.1088/0305-4470/20/11/034}.
\bibitem{Kobe} Kobe D.H., Liang J.Q., Phys. Rev. A, 1988, \textbf{37}, No.~4, 1133--1140, \doi{10.1103/PhysRevA.37.1133}.
\bibitem{ShekaPRA} Sheka  D.D., Mertens  F.G., Phys. Rev. A, 2006, \textbf{74}, 052703 (5 pages),
         \doi{10.1103/PhysRevA.74.052703}.
\bibitem{Rastelli} Rastelli G., Houzet M., Glazman L., Pistolesi F., C. R. Phys., 2012, \textbf{13}, No.~5, 410--425, \\ \doi{10.1016/j.crhy.2012.03.001}.
\bibitem{Skorobagatko} Skorobagatko G.A., Condens. Matter Phys., 2018, \textbf{21}, No.~2, 23703 (8 pages), \doi{10.5488/CMP.21.23703}.
\bibitem{Shekhter} Shekhter R.I., Gorelik L.Y., Glazman L.I., Jonson M., Phys. Rev. Lett., 2006, \textbf{97}, No.~15, 156801, \\ \doi{10.1103/PhysRevLett.97.156801}.
\bibitem{Mertens} Ivanov B.A., Schnitzer H.J., Mertens F.G., Wysin G.M., Phys. Rev. B, 1998, \textbf{58}, No.~13,
                 8464--8474, \\ \doi{10.1103/PhysRevB.58.8464}.
\bibitem{Sheka} Sheka  D.D., Yastremsky I.A., Ivanov B.A., Wysin G.M., Mertens F.G., Phys. Rev. B, 2004, \textbf{69}, 054429 (13~pages),
         \doi{10.1103/PhysRevB.69.054429}.
\bibitem{ShekaMer01}  Sheka D.D., Ivanov B.A., Mertens F.G., Phys. Rev. B, 2001, \textbf{64}, 024432 (15 pages), \\ \doi{10.1103/PhysRevB.64.024432}.
\bibitem{Nagaosa} Iwasaki J., Beekman A.J., Nagaosa  N., Phys. Rev. B, 2014, \textbf{89}, 064412 (7 pages), \\ \doi{10.1103/PhysRevB.89.064412}.
\bibitem{IvWysinPRB02} Ivanov B.A., Wysin G.M., Phys. Rev. B, 2002, \textbf{65}, 134434 (17 pages), \doi{10.1103/PhysRevB.65.134434}.
\bibitem{Uzunova19} Uzunova V.A., Ivanov B.A., Low Temp. Phys., 2019, \textbf{45}, 92--97, \doi{10.1063/1.5082327}, [Fiz. Nizk.
                   Temp., 2015, \textbf{45}, No.~1, 104--110 (in Russian)].
\bibitem{Garst} Sch\"{u}tte C., Garst M., Phys. Rev. B, 2014, \textbf{90}, 094423 (13 pages), \doi{10.1103/PhysRevB.90.094423}.
\bibitem{GarstFNT} Schroeter S., Garst M., Low Temp. Phys., 2015, \textbf{41}, 817 (10 pages), \doi{10.1063/1.4932356},
                   [Fiz. Nizk. Temp., 2015, \textbf{41}, No.~10, 1043--1053].
\bibitem{deGennes} De Gennes P.G., Superconductivity of Metals and Alloys, CRC Press, Boca Raton, 1999, \\ \doi{10.1201/9780429497032},
                   [Mir, Moscow, 1968 (in Russian)].
\bibitem{Landau9} Lifshitz E.M., Pitaevskii  L.P., Statistical Physics, Part 2: Theory of the Condensed State, Vol. 9, Oxford,
                   1980, [Nauka, Moscow, 1978 (in Russian)].
\bibitem{Bogolyubov} Bogolyubov N.N. (Jr.), A Method for Studying Model Hamiltonians, Pergamon Press, Oxford, New York, 1972, \\ \doi{10.1016/C2013-0-02470-X},
                     [Nauka, Moscow, 1974 (in Russian)].
\bibitem{Ivanov95} Ivanov B.A., JETP Lett., 1995, \textbf{61}, No.~11, 917--920, [Pis'ma Zh. \'{E}ksp. Teor. Fiz., 1995, \textbf{61}, 898--902
          (in Russian)].
\bibitem{IvShekaJETPL} Ivanov B.A., Sheka D.D., JETP Lett., 2005, \textbf{82}, No.~7, 489--493, \doi{10.1134/1.2142872},
         [Pis'ma Zh. \'{E}ksp. Teor. Fiz., 2005, \textbf{82}, No.~7, 489--493 (in Russian)].


\end{thebibliography}
\end{document}